\documentclass[prb,aps,twocolumn]{revtex4}
\usepackage{amsfonts,amsmath,amssymb,latexsym}
\usepackage{epsfig}
\usepackage{graphics}

\begin{document}

\title{ Unusual broadening of wave packets on lattices }
\author {K. Sch\"onhammer}
\affiliation{Institut f\"ur Theoretische Physik, Universit\"at
  G\"ottingen, Friedrich-Hund-Platz 1, D-37077 G\"ottingen}

\date{\today}

\begin{abstract}

  The broadening of one-dimensional
  Gaussian wave packets is presented in all textbooks
on quantum mechanics. It is used as an example to elucidate
Heisenberg's uncertainty relation.
The behaviour on a lattice is drastically different if the 
amplitude and (or) phase 
of the wave packet varies on the scale of the lattice constant.
Important examples are  very narrow wavepackets or wavepackets with an
average velocity
 comparable to the maximum velocity on the lattice.
Analytical and numerical results for the time dependence of
wave packets on a one-dimensional lattice are presented. The long-time
limit of the shape of the wave packet is discussed.

\end{abstract}

\maketitle

\section{Introduction}

Even before treating
how the probability amplitudes in quantum mechanics vary in 
continuous space
  Richard Feynman \cite{Feynman} 
in volume III of his famous ``Lectures on Physics''
 addresses what happens 
if one puts a single electron on a line of atoms.
He arrives at the one-dimensional time independent
Sch\"odinger equation by studying the limit where the lattice 
constant $a$ goes to zero.
 The one-dimensional model 
studied by Feynman is usually called the ``nearest neighbour tight-binding
model'' and is treated in solid state physics textbooks.\cite{AM}
More recently in the field of quantum computing this model is called
a ``continuous time quantum walk''. \cite{Farhi} Here the solid state physics
point of view is taken.

Feynman discusses the dynamics of wave packets on
the lattice and shows that if they have a predominant wave number $k_0$
they move through the lattice with the
group velocity $d\omega/dk$. 
In a footnote he adds ``Provided we do
not try to make the packet too narrow''. It is one goal
of this paper to elucidate what happens in this case.
We examine the effects which occur 
 for ``too narrow'' wave packets.  They are 
 drastically different from the
 behaviour  in continuous space. The second important difference between
 the lattice case and continuous space concerns the
time dependence of the shape
 of the wave packet on its average velocity.
As discussed in section II, on
one-dimensional lattices the wave number $k$ of a plane wave state
is restricted to the interval $-\pi/a$ to $\pi/a$, were $a$ is the
lattice constant and the corresponding velocity has a maximum value
in this first Brilloin zone. \cite{AM} 
This maximum velocity e.g. largely determines the time dependence of
the wavepacket when the electron initially is put on a single lattice
site. This is similar to the spreading of
the information about a local perturbation 
in spin systems with finite range interactions.
Lieb and Robinson \cite{LiebRobinson}
showed that for such systems
a finite bound for the group velocity exist, with which
the information 
propagates in the system.
The topic of Lieb-Robinson bounds has become
a useful tool when studying nonequlibrium phenomena in
 quantum many body systems. \cite{Everts}
 The behaviour of  a wave packet of a single
electron initially localized in a finite region of the lattice,
 is a simple example of a similar physics.

Various aspects of the dynamics of wave packets were treated in this
journal for the case of a one dimensional continuum \cite{Mita} but
not on lattices.

The paper is structured as follows.
In section II we present the basic concepts of the quantum mechanics
of a particle on a one-dimensional lattice. The corresponding dynamics
is addressed in section III in the Schr\"odinger as well as 
the Heisenberg picture. It is shown that on the
lattice the
uncertainty relation for the position and the velocity differs
from the continuum case. The uncertainty product can even be zero
in the lattice case as the position operator has a discrete spectrum.

Numerical results for the dynamics of narrow and more extended wave packets are
presented in section IV. If the initial state is confined to one
or a few lattice sites the results differ drastically from
the continuum case discussed in textbooks. For Gaussian
initial states of sufficiently large width the behaviour 
on the continuum is recovered for small average velocity. 
For average velocities of the order of the maximum velocity 
again large differences to the continuum case occur. 

In the discussion of the long-time limit the 
probability distribution of the velocity in the initial state
plays the central role.

\section{One-dimensional lattice}

An infinite one-dimensional lattice is considered with lattice sites
at positions $na$, were $a$ is the lattice constant and
the integer $n$ runs
from $-\infty$ to $\infty$. The state for the particle at position $an$
is denoted by $|n\rangle$. The states are assumed to be orthonormal
and to span the infinite Hilbert space
\begin{equation}
\label{orthonormal}
 \langle m|n\rangle= \delta_{mn},~~~\sum_n|n\rangle \langle n|=\hat 1.
\end{equation} 
A general normalized state for the particle on the lattice takes the
form
\begin{equation}
\label{state}
|\phi\rangle= \sum_n c_n|n\rangle~,
\end{equation} 
where the complex weights $c_n$ fulfill $\sum_n|c_n|^2=1$. The
$c_n=\langle n|\phi \rangle$ play the role of the wave function
$\phi(x)=\langle x|\phi \rangle$ in the continuum at position $x=na$.

The position operator $\hat x$ on the lattice is defined as
\begin{equation}
\label{xoperator}
 \hat x=\sum_n|n\rangle na\langle n|.
\end{equation} 
It is useful to introduce translation operators $\hat T_{\pm a}$ by one
lattice spacing
\begin{equation}
\label{Toperators}
\hat T_{\pm a}|n\rangle=|n\pm 1\rangle~.
\end{equation} 
Multiplying from the right by $\langle n|$ yields after summation over
$n$ using Eq. (\ref{orthonormal})
\begin{equation}
\hat T_{\pm a}=\sum_n|n\pm 1\rangle \langle n|~.
\end{equation} 
As $\hat T_a\hat T_{-a}|n\rangle=\hat T_{-a}\hat T_a|n \rangle$ the operators $\hat T_a$ and
$\hat T_{-a}$ commute and one easily shows that $\hat T_{\pm a}^\dagger=T_{\mp
  a}$ holds.

The eigenstates of the unitary operators $\hat T_{\pm a}$ are plane wave
 states $|k\rangle$
\begin{equation}
\label{planewave}
|k\rangle=c\sum_ne^{ikan}|n\rangle~, 
\end{equation} 
with a constant $c$ to be fixed later. As
$e^{i(k+2\pi m/a)an}=e^{ikan}$ for all integers $m$ the complete set of
eigenstates $|k\rangle$ is obtained by restricting the $k$ values to 
the interval from $-\pi/a$ to  $\pi/a$, in solid state physics
called the first Brillouin zone. In order to have the
completeness relation 
 \begin{equation}
\label{kcomplete}
\int_{-\pi/a}^{\pi/a}|k\rangle \langle k|dk=\hat 1,
\end{equation} 
the constant $c$ has to be properly fixed. Assuming this completeness
relation to hold, one obtains
\begin{equation}
1=\langle n|n\rangle=\int_{-\pi/a}^{\pi/a}\langle n|k\rangle \langle
k|n\rangle dk=|c|^22\pi/a.
\end{equation} 
This implies $c=\sqrt{(a/(2\pi)}$, if $c$ is chosen as a
positive real number. The $k$-components of $|\phi \rangle$ are therefore given 
by 
 \begin{equation}
\label{phivonk}
\langle k|\phi \rangle =\sqrt{\frac{a}{2\pi}}\sum_ne^{-ikan}c_n~.
\end{equation} 
This is a $(2\pi/a)$-periodic function of the wave number $k$.

The eigenvalues of the translation operators  $\hat T_{\pm a}$ can easily
be read off Eq. (\ref{planewave}) using the definition
Eq. (\ref{Toperators})
 \begin{equation}
\hat T_{\pm a}|k\rangle=e^{\mp ika}|k\rangle~.
\end{equation} 
With the wave number operator $\hat k$
 \begin{equation}
\label{koperator}
\hat k=\int_{-\pi/a}^{\pi/a}|k\rangle k \langle k|dk
\end{equation}  
the translation operators therefore take the form
 \begin{equation}
\label{Tofk}
\hat T_{\pm a}=e^{\mp i\hat ka}~.
\end{equation}
The operator $\hbar \hat k$ is called the quasi-momentum operator.
  
The position operator has simple commutation relations with
the translation operators. For an arbitrary basis state $|n\rangle$
one has
 \begin{equation}
(\hat x \hat T_{\pm a}-\hat T_{\pm a}\hat x)|n\rangle=a(n\pm 1-n)|n\pm 1\rangle
=\pm a\hat T_{\pm a}|n\rangle~,
\end{equation} 
which implies
 \begin{equation}
\label{xcomm}
[\hat x,\hat T_{\pm a}]=\pm a \hat T_{\pm a}~.
\end{equation} 
This result will be used in Section III.\\

For the Hamiltonian which allows the particle to propagate in
the lattice by hopping via
neighbouring sites we use
identical real hopping matrix elements $\epsilon_{01}$ between all
 neighbouring sites
\begin{eqnarray}
\label{Hamiltonian}
\hat H&=&\epsilon_0 \sum_n|n\rangle \langle n|+\epsilon_{01}
 \sum_n(|n+1\rangle \langle n|+|n\rangle \langle n+1|), \nonumber \\
&=& \epsilon_0 \hat 1 +\epsilon_{01}(\hat T_a+\hat T_a^\dagger)~.
\end{eqnarray}
The value of the site energy $\epsilon_0$ is irrelevant for the
broadening of wave packets.

As $\hat H$ depends linearly on the translation operators $\hat T_{\pm a}$ the
states $|k\rangle$ are also the eigenstates of $\hat H$
 \begin{equation}
\label{dispersion}
\hat H|k\rangle=\epsilon_k|k\rangle,~~~
   \epsilon_k=\epsilon_0 +2\epsilon_{01}\cos(ka)~.
\end{equation}
The energy eigenvalues $\epsilon_k$ lie in a band of
(total) bandwidth $B=4 |\epsilon_{01}|$.
If one defines an effective mass $m_{\rm eff} $ via
 \begin{equation}
\label{effectivem}
\epsilon_{01}=-\frac{\hbar^2}{2m_{\rm eff}a^2} 
\end{equation}
the energy dispersion for $|ka|\ll 1$ is given by
 \begin{equation}
 \epsilon_k=V_0+\frac{\hbar^2k^2}{2m_{\rm eff}}+O(k^4a^4)
\end{equation}
with $V_0=\epsilon_0+2\epsilon_{01}$.
If one takes the limit $a\to 0$ for fixed  $m_{\rm eff} $ the
nonrelativistic energy dispersion of a particle on a continuous one
dimensional line (in a constant potential $V_0$) is obtained.
This can also be seen in the site representation
 \begin{equation}
\langle n |\hat H|\phi \rangle= V_0c_n-\frac{\hbar^2}{2m_{\rm eff}} 
(c_{n+1}-2c_n+c_{n-1})/a^2~.
\end{equation}
 The parenthesis devided by $a^2$ goes over to the second
 derivative in the continuum limit, leading to 
 the well known kinetic energy term of the
 particle.

\section{Dynamics of wave packets on a one-dimensional lattice}

In this section we present general results for the dynamics of
 wave packets on lattices. The wave packet at time $t=0$ is of the
general form of Eq. (\ref{state}) and assumed to be located near the
the origin ($n=0$). To study the time dependence we first present the
solution of the time dependent Schr\"odinger equation. To address
directly the time dependence of the width of the wave packet we 
also use the Heisenberg picture. As Heisenberg's
uncertainty relation is useful for the discussion
of the width on the continuum, its state dependent  form 
for the lattice is discussed.

As it only would lead to irrelevant 
phase factors we put $\epsilon_0=0$ in the following.

\subsection{Schr\"odinger picture}

If one takes the state in Eq. (\ref{state}) as the initial state at
time $t=0$, the coefficients at $t>0$ become time dependent:
$c_n \to c_n(t)$. As the Hamiltonian is time independent the formal 
result for the  $c_n(t)$ is straightforward
\begin{equation}
c_n(t)=\langle n|\phi(t)\rangle=\sum_mc_m
\langle n|e^{-i\hat Ht/\hbar}|m\rangle 
\end{equation}
The calculation of the matrix elements of the evolution operator in the site
representation can be reduced to an integration 
 using the completeness relation Eq. (\ref{kcomplete})
\begin{eqnarray}
\label{matrixelements}
\langle n|e^{-i\hat Ht/\hbar}|m\rangle &=&  
\int_{-\pi/a}^{\pi/a}\langle m|k\rangle \langle k|n\rangle e^{-i\epsilon_kt/\hbar}dk
  \\
&=& \frac{a}{2\pi}\int_{-\pi/a}^{\pi/a}e^{ika(m-n)}
e^{-i\epsilon_kt/\hbar}dk ~.\nonumber
\end{eqnarray}
The second line explicitely shows  the discrete translational invariance
\begin{equation}
\langle n|e^{-i\hat Ht/\hbar}|m\rangle=\langle n-m|e^{-i\hat Ht/\hbar}|0\rangle~.
\end{equation}
As $\epsilon_k$ is an even function of $k$ the factor $e^{ika(m-n)}$
in Eq. (\ref{matrixelements}) can be replaced by $\cos[ka(m-n)]$.
Switching to dimensionless units $\tilde k=ka$ and $\tilde t
=-2\epsilon_{01}t/\hbar$ one obtains
\begin{equation}
\label{cBessel}
\langle n|e^{-i\hat Ht/\hbar}|0\rangle=\frac{1}{\pi}
\int_0^\pi \cos(n\tilde k)e^{i\tilde t
  \cos\tilde k}d\tilde k \equiv g_n(t)~.
\end{equation}
As shown in the appendix 
the function $ g_n(t)$ is real for even values of $n$ and purely
imaginary for odd values of $n$. As it is an even function of $n$
it is sufficient to consider values $n\ge 0$.
 Apart from an additional factor $i^n$,  $ g_{\pm n}(t)$
equals the Bessel function 
of integer order $J_n(\tilde t)$. This result was discussed earlier
in the context of continuous time quantum walks. \cite{Blumen}

If the characterization of the initial state $|\phi \rangle$ is 
presented in terms of the wave number amplitudes $\langle k|\phi
\rangle$ a more direct way to calculate $c_n(t)$ is
to  expand $|\phi(t)\rangle$ in terms of the $k$-states 
\begin{equation}
\langle n|\phi(t)\rangle=\sqrt{\frac{a}{2\pi}}\int_{-\pi/a}^{\pi/a}
e^{ikan}e^{-i\epsilon_kt/\hbar}\langle k|\phi \rangle dk~.
\end{equation} 
If $|\langle k|\phi\rangle|$ is strongly peaked around a value $k=k_0$
the integral can be performed by expanding $\epsilon_k$ 
(for arbitrary dispersion) around $k_0$. Then the average velocity
of the wave packet
is determined by the group velocity $v_k=(d\epsilon_k/dk)/\hbar$
at $k=k_0$, as discussed by Feynman.\cite{Feynman}

 In section IV results
for the time dependence of wave packets will be presented by 
numerically calculating the integral in Eq. (\ref{cBessel}) 
as a function of $\tilde t$ and $n$.\\

\subsection{Heisenberg picture}

For an analytical understanding of the average position and the width of the 
wave packet it is useful to switch to the Heisenberg picture. For a
general observable described by the time independent operator $\hat A$  the 
time dependent expectation value is given by
\begin{equation}
\langle \hat A  \rangle_t = 
\langle \phi |e^{i\hat Ht/\hbar}\hat  A e^{-i\hat Ht/\hbar}|\phi \rangle
= \langle \phi |\hat A_H(t)|\phi \rangle~.
\end{equation}
 Taking the time derivative of the Heisenberg operator $\hat A_H(t) $
yields the equation of motion in the form
\begin{equation}
i \hbar \frac{d}{dt}\hat A_H(t)=e^{i\hat Ht/\hbar} [\hat A,\hat H]
e^{-i\hat Ht/\hbar}.
\end{equation}
For $\hat A=\hat x$ the commutator $[\hat x,\hat H]$ is proportional to
 $[\hat x,\hat T_a+\hat T_{-a}]$. With Eq. (\ref{xcomm}) one obtains
\begin{equation}
\label{Hvelocity}
 \frac{d\hat x_H(t)}{dt}=\hat v_H(t)~,
\end{equation}
with the velocity operator 
\begin{equation}
\label{velocityop}
 \hat v=\frac{1}{i\hbar}a\epsilon_{01}(\hat T_a-\hat T_{-a})
=-2a\epsilon_{01}\sin(\hat k a)/\hbar
\end{equation}
In the position representation one has with Eq. (\ref{effectivem})
\begin{equation} 
\langle n|m_{\rm eff}\hat v|\phi \rangle= \frac{\hbar}{i} 
\frac{c_{n+1}-c_{n-1}}{2a}~.
\end{equation}
In the continuum limit this leads to the familiar result
$\langle x|\hat p|\phi
\rangle =(\hbar/i)\phi '(x)$. 

The  eigenstates of $\hat v$ are the plane wave states $|k\rangle$
\begin{equation}
\hat v|k\rangle =v_k|k\rangle,~~~
v_k=v_{\rm max}\sin(ka), 
\end{equation}
with
\begin{equation}
v_{\rm max}=-2a\epsilon_{01}/\hbar= \hbar/(am_{\rm eff})
\end{equation}
Note that $v_k=(d\epsilon_k/dk)/\hbar$ for
$\epsilon_{10}=-|\epsilon_{10}|$
has its maximal value $2|\epsilon_{10}|a/\hbar$  at $k=\pi/(2a)$.

As $\hat T_a$ and $\hat T_{-a}$ commute with each
 other $T_{\pm a, H}(t)=\hat T_{\pm a}$
for the Hamiltonian $\hat 
H$ in Eq. (\ref{Hamiltonian}).
This implies the time independent expectation value
\begin{equation}
\label{expTa}
\langle \hat T_{\pm a}\rangle_t = \sum_n c_{n\pm 1}^*c_n~.
\end{equation}

 $T_{\pm a, H}(t)=\hat T_{\pm a}$ also leads to
$\hat v_H(t)=\hat v$. 
Therefore
Eq. (\ref{Hvelocity}) can be
trivially integrated

\begin{equation}
\label{x1}
\hat x_H(t)=\hat x +\hat v t~.
\end{equation}
Squaring leads to
\begin{equation}
\label{x2}
\hat x^2_H(t)=\hat x^2 +(\hat x\hat v+\hat v \hat x) t+\hat v^2t^2~.
\end{equation}
The last two equations have
 the same form as for a particle on a continuous line. In that
case one has the familiar result that the
velocity operator $\hat v$ is given by the momentum operator $\hat p$ devided
by the mass $m$ of the particle.
Using   $\hat v$ instead of $\hat p$ the Born-Heisenberg commutation relation
reads $[\hat x,\hat v]=(i\hbar/m)\hat 1$. 
For the lattice Hamiltonian in Eq. (\ref{Hamiltonian})
 the velocity operator is 
given by the expression in Eq. (\ref{velocityop}). 
Using Eq. (\ref{xcomm}) one sees that the commutator for the lattice
is not proportional to the unit operator
\begin{eqnarray}
\label{latticecomm}
[\hat x,\hat v]&=&\frac{a \epsilon_{01}}{i\hbar}[\hat x,\hat T_a-\hat T_{-a}]
                   =\frac{a^2 \epsilon_{01}}{i\hbar}(\hat T_a+\hat T_{-a})~.\\
  \nonumber
  &=& \frac{i\hbar}{m_{\rm eff}} (\hat T_a+\hat T_{-a})/2~.
\end{eqnarray}
For a Hamiltonian with hopping matrix elements not only to nearest
neighbours additional terms appear in the result for $ [\hat x,\hat v]
$.
The commutator of $\hat x$ and the quasi-momentum operator $\hbar \hat
k$ is also not proportional to the unit operator. \cite{KS} Here the
focus is on the velocity operator
defined as the time derivative of the position operator.
In Eq. (\ref{x2}) it enters
the description of the broadening of the wavepacket.

For the general intial state $|\phi \rangle$  in Eq. (\ref{state})
the expectation value of
the commutator is given by
 \begin{equation}
\label{uc}
\langle[\hat x,  \hat v]\rangle=\frac{i\hbar}{m_{\rm eff}}
 {\rm Re} \sum_n c_{n+1}^*c_n~.
\end{equation}
For a single site initial state $|\phi\rangle =|m\rangle$
this expectation value vanishes.  This is in sharp contrast
to a very broad and smooth
initial state with $c_{n+1}\approx c_n$ where it approaches
the continuum value $i\hbar/m_{\rm eff}$.

Equations (\ref{x1}) and  (\ref{x2}) allow to express the time dependence
  of the average position $\langle \hat x \rangle_t$ and the width
 \begin{equation}
(\Delta x)_t=\sqrt{\langle \hat x^2 \rangle_t-\langle \hat x \rangle_t^2  }
\end{equation}
   of the 
wave packet in terms of the initial state expectation values of the operators
$\hat x$ and $\hat v$ and the products of them appearing in Eq.  (\ref{x2}).
For the average of $\hat v$ and $\hat v^2$ one obtains

\begin{eqnarray}
\label{vvoncn}
\langle \hat v \rangle
  &=&-v_{\rm max} {\rm Im}\sum_nc_{n+1}^*c_n,\\
  \label{v2voncn}
\langle \hat v^2 \rangle 
  &=& \frac{v_{\rm max}^2}{2
      }
   \left (1-{\rm Re}\sum_nc_{n+2}^*c_n\right)~.
\end{eqnarray}

Also required to obtain $(\Delta x)_t$ is the expectation value
 \begin{equation}
\label{xv}
\langle \hat x \hat v  \rangle+ \langle \hat v \hat x  \rangle
 =-av_{\rm max} {\rm Im}\sum_n(2n+1)c_{n+1}^*c_n
\end{equation}
For the initial states treated in the following
$\langle \hat x \hat v  \rangle+ \langle \hat v \hat x  \rangle
-2\langle \hat x \rangle \langle \hat v \rangle $
vanishes and the time dependence of the width of the wave packet is
given by
\begin{equation}
  (\Delta x)_t=\sqrt{(\Delta x)^2+t^2 (\Delta v)^2 }~.
 \end{equation}  
 The product of the initial state uncertainties obeys
 the uncertainty relation \cite{Merzbacher}
 
\begin{equation}
  \Delta x \Delta v\ge \frac{1}{2}|\langle [\hat x ,\hat v]\rangle|~.
 \end{equation}

For the discussion of the results in section IV it is 
helpful to also consider the (time-independent) probability distribution
of the velocity
 \begin{equation}
\label{pvonv}
p_{\rm vel}(v)=\langle \delta (v-\hat v)\rangle
=\int_{-\pi/a}^{\pi/a}|\langle k|\phi \rangle|^2 \delta(v-v_{\rm max}\sin{ka})dk~.
\end{equation}
 The integral
can be performed using the general formula
\begin{equation}
 \int f(k) \delta(g(k))dk=\sum_i \frac{f(k_i)}{|g'(k_i)|}~,
\end{equation}
where the $k_i$ are the positions of the zeros of $g(k)$ in the integration
range. For $|v|<v_{\rm max}$ there are two zeros $ak_1(v)=\arcsin(v/v_{\rm max})$
and $ak_2(v)={\rm sign}(v)\pi-ak_1(v)$, implying
$\cos ak_1(v)=-\cos ak_2(v)$.
 This yields with
$|\cos(k_i(v))|=\sqrt{1-(v/v_{\rm max})^2}$
\begin{equation}
\label{pvonve}
p_{\rm vel}(v)=\frac{1}{a v_{\rm max}
  \sqrt{1-(v/v_{\rm max})^2}}(|\langle k_1(v)|\phi\rangle|^2+
|\langle k_2(v)|\phi\rangle|^2)~,
\end{equation}
and $p_{\rm vel}(v) $ vanishes for $|v|>v_{\rm max}$. The explicit
expressions for $p_{\rm vel}(v)$ presented in section IV hold for
$|v|\le v_{\rm max}$.

A look at the higher moments $\langle \hat x^n_H(t)\rangle$ in the long time
limit turns out to be useful. For the localized initial states $|\phi \rangle$
considered
in this paper they all exist and $\hat x_H(t)=\hat x +\hat v t$ implies
for all integers $n$
\begin{equation}
 \label{scaledmoments}
\lim_{t \to \infty}\langle (\hat x_H(t)/t)^n\rangle=\langle \hat v^n\rangle~,
\end{equation}
i.e. the {\it continuous} probability distribution
$ p_{\rm vel} $ determines all scaled moments in the long
time limit. At arbitrary time $t$ 
the moments $\langle \hat x^n_H(t)\rangle $ are determined
by the {\it discrete}
probabilities $|c_m(t)|^2$. As discussed in section IV
the convergence of the probability distributions is more delicate than
that of the moments.

\subsection{Shape dependence of the wave packet on the average
  velocity}

In this subsection we compare the time dependence of
the shape of the wave packets for the initial state $|\phi \rangle$
in Eq. (\ref{state}) and the ``boosted'' state
\begin{equation}
\label{boosted}
  |\phi\rangle^{(k_0)}=e^{ik_0 \hat x}|\phi \rangle~,
\end{equation}
with $k_0$ a real number. The corresponding expansion coefficients are
given by  $c_n^{(k_0)}=e^{ik_0 a n}c_n$.
Using Eq. (\ref{x1}) the time dependent boosted state is given by
\begin{eqnarray}
  |\phi (t)\rangle^{(k_0)}&=& e^{-iHt/\hbar}e^{ik_0 \hat x}|\phi \rangle
       =e^{ik_0\hat x_H(t)}|\phi(t)\rangle \\ \nonumber
  &=& e^{ik_0(\hat x+\hat v t)}|\phi(t)\rangle~.
\end{eqnarray}
The prefactor of the form $e^{\hat A +\hat B}$ is difficult to simplify
in the lattice case. This is different for the continuum where
 the Baker-Haussdorf formula \cite{Merzbacher} can be used. It reads
\begin{equation}
 \label{BH}
e^{\hat A +\hat B} =e^{\hat A}e^{\hat B}e^{-\frac{1}{2}[\hat A,\hat B]},~~\mbox{if}~~
[\hat A,[\hat A,\hat B]]=0=[\hat B,[\hat A,\hat B]].
\end{equation}
As $[\hat x,\hat v]$ is proportional to the unit operator
in the continuum case 
the condition is fulfilled and with
$v_0=\hbar k_0/m$ it reads
\begin{equation}
  e^{ik_0(\hat x+\hat v t)}  =e^{ik_0 \hat x}e^{iv_0t\hat p/\hbar}e^{ik_0^2\hbar t/(2m)}~.
\end{equation}
The second factor is the translation operator \cite{Merzbacher}
by the distance $v_0t$.
Applying it to the position state $\langle x| $ to the left yields
$\langle x| e^{iv_0t\hat p/\hbar}=\langle x-v_0t|  $. Squaring leads
to the important result
\begin{equation}
|\langle x|\phi(t) \rangle^{(k_0)}|^2=|\langle x-v_0 t|\phi(t) \rangle|^2~.
 \end{equation} 
The boost does not change the shape of the time dependent wave packet.
It only gets the time dependent shift as expected from Galilean invariance
in the continuum case. This well known result for the Gaussian case
holds for arbitrary shapes of $|\phi \rangle$.

As the commutator $[\hat x,\hat v]$ is {\it not}
proportional to the unit operator in the lattice case 
(see Eq. (\ref{latticecomm}))
this shape independence does {\it not} hold on the lattice as confirmed by the 
numerical results in section IV. 
As no simple
result for $  e^{ik_0(\hat x+\hat v t)} $ can be obtained 
we analytically only show in section IV.B (see Eq. (\ref{vuncert})) that
the
time independent velocity uncertainty depends on $k_0$.

\section{Numerical results and their interpretation}

As mentioned in the introduction the broadening of wavepackets on a
lattice differs strongly from the continuum case discussed in quantum
mechanics textbooks when in the coefficients $c_n=\tilde c_ne^{i\phi_n}$,
with $\tilde c_n$ and $\phi_n$ real, the  $\tilde c_n$ and (or)   
 $\phi_n$ vary rapidly on the scale of the lattice distance $a$.
In subsection A the phases are put to zero, implying zero average
velocity of the packet. The case of nonzero phases is discussed in
subsection B.

In this section we work with dimensionless units 
$\hbar=1$ and $a=1$. In order for $\tilde t=t$ to hold
we choose $\epsilon_{10}=-1/2$, which implies
\begin{equation}
\epsilon_k=-\cos k~,~~~~v_k=\sin k ~,
\end{equation}
i.e. $v_{\rm max}=1$. The small $k$ curvature of $ \epsilon_k$ 
corresponds to an effective mass $m_{\rm eff} =1$.

\subsection{Initial state with real expansion coefficients $c_n$}
In this subsection we discuss wave packets with real coefficients $c_n$
in Eq. (\ref{state}).
According to Eq. (\ref{vvoncn}) they have zero average velocity.

We start by discussing a {\it single site} initial state.
Because of the discrete translational invariance we can choose
 $|\phi \rangle =|0\rangle$,
i.e the initial state is localized at the origin.
As $\langle k|0\rangle=1/\sqrt{2\pi}$
 all $k$ states contribute equally. One can therefore expect
the deviation from the behaviour of wave packets in
the continuum to be strongest for a single site initial state.

Equation ({\ref{v2voncn}) leads to
 $\langle \hat v ^2\rangle=1/2$. As $(\Delta x)_{t=0}=0$
 the width of the wave packet increases linearly with
time $(\Delta x)_t=t/\sqrt{2}$ and the  uncertainty product is 
 given by  $(\Delta x)_t \Delta v =t/2$.

 The probability to find the particle at site $|n\rangle$ at time
$t$ is given by (see the discussion following Eq. (\ref{cBessel}))
\begin{equation}
 |c_n(t)|^2=|\langle n|e^{-i\hat Ht/\hbar}|0\rangle|^2
=J^2_n(t)~.
\end{equation}
In Fig.1 we show the probability $|c_n(t)|^2$ as a function of $n$
for  two different times.
\begin{figure}[h]
\label{fighc}
\centering
\epsfig{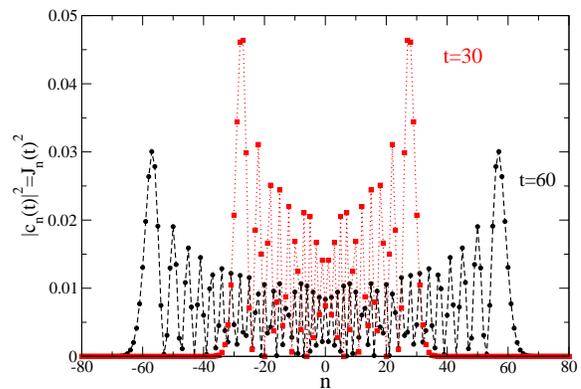}
\vspace{0.5cm}
\caption{Probability to find the particle at site $n$
if it was localized in the state $|0\rangle$ at $t=0$ for two different times:
t=30,  squares on the dotted line; t=60,  circles on the dashed line.}
\end{figure}
We have connected the discrete points at integer $n$ by a dotted line
 for $t=30$ and a dashed line for $t=60$. The result is drastically
different from the broadening of a wave packet in the continuum.
The dominant peaks indicate that
 the particle has a high probability to move with the
maximum velocity in both directions.
 This can be understood using the probability
distribution of the velocities in Eq. (\ref{pvonve}). 
For $|\phi \rangle=|0\rangle$ one has $ |\langle k_i(v)|\phi\rangle|^2=
1/2\pi$ leading 
\begin{equation}
\label{pv0}
p_{\rm vel}(v)=\frac{1}{\pi}\frac{1}{\sqrt{1-v^2}}~.
\end{equation}
The divergence of this function when approaching $\pm v_{\rm max}=\pm 1$
is responsible for the large weight near $n\approx \pm t$ 
in Fig. 1. The approach of the discrete probability density $|c_n(t)|^2$
to the continuous distribution  $p_{\rm vel}(n/t)/t  $
one would expect from
Eq. (\ref{scaledmoments}) is obviously not a smooth one . \cite{weaklimit}
A simple averaging procedure better shows the importance of
$p_{\rm vel}(n/t)/t  $ for large enough $t$.
We define
\begin{equation}
p_{\rm av}(n,t;n_0)=\frac{1}{2n_0+1}\sum_{j=-n_0}^{n_0}|c_{n+j}(t)|^2~.  
\end{equation}
In Fig. 2 the continuous $p_{\rm vel}(n/t)/t  $
is compared to the discrete averaged probability
$ p_{\rm av}(n,t;n_0) $ for $t=60$
and $n_0=3$.

\begin{figure}[h]
\label{figaverage}
\centering
\epsfig{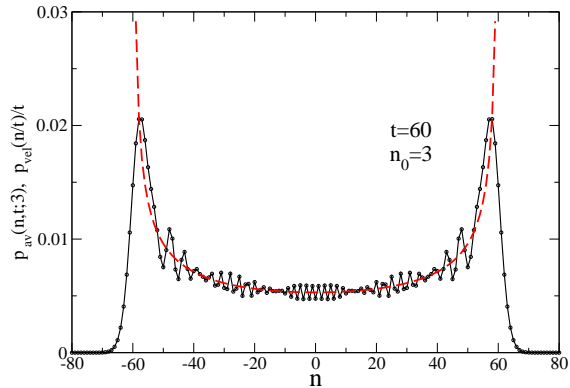}
\vspace{0.5cm}
\caption{ Averaged probability distribution
  $ p_{\rm av}(n,t;n_0) $ for $t=60$
and $n_0=3$ for the initial state $|0\rangle$. 
  The dashed line shows  $p_{\rm vel}(n/t)/t  $. }
\end{figure}

To further elucidate the rather well defined edges of the packet
in Fig. 1 at $n\approx \pm t$ we show the probability
  $|c_n(t)|^2$ for $n=30$ and 
$n=60$ as a function of time in Fig. 3.

\begin{figure}[h]
  \label{Bessel}
  \centering
\epsfig{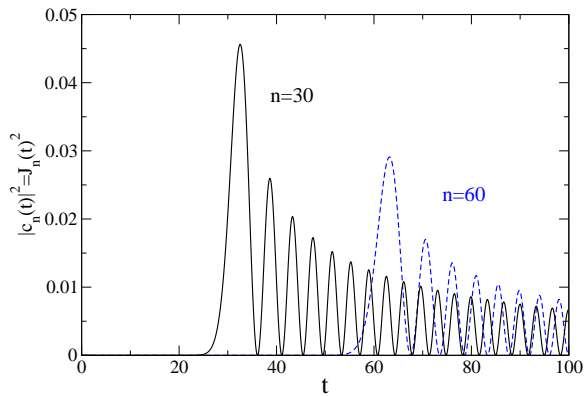}
\vspace{0.5cm}
\caption{Probability to find the particle at two different sites
as a function of time time $t$,
if it was localized in the state $|0\rangle$ at $t=0$: $n=30$ (full line)
and  $n=60$ (dashed line).}
\end{figure}
The short time behaviour of $g_n(t)$ is discussed in the
appendix (see Eq. (\ref{shorttime})).
This shows that  $|c_n(t)|^2=J_n^2(t) $,
is strongly suppressed for times $t$ small compared to $n$. 
 Therefore the information that the particle was located at the
origin at time $t=0$ arrives at site $n$  not before the time 
determined by 
$tv_{\rm max}=an$, which for  $v_{\rm max}=1$ and $a=1$ is given
by $t=n$. This is similar to the spread of information in spin systems
on a lattice with finite range interactions. \cite{LiebRobinson,Everts}

 For times $t\gg n$ the probability   $|c_n(t)|^2=J_n^2(t) $
 decays like $1/t$ in an oscillatory manner
 \begin{eqnarray}
   J_n^2(t) &\to& \frac{2}{\pi t} \cos^2(t-\pi/4), ~{\rm for} ~n
                  ~~{\rm even} \\  
          &\to& \frac{2}{\pi t} \sin^2(t-\pi/4), ~{\rm for}~ n ~~{\rm odd}~.
\end{eqnarray}              
This follows from
the large time behaviour of the Bessel functions of integer order
discussed in the appendix. It describes the oscillatory behaviour
 of  $|c_n(t)|^2$ very close to the origin in Fig. 1. For $t=n_t\pi$ with
 $n_t$ integer these small
 $n$ oscillations are suppressed and $ |c_n(t)|^2 \to
 1/(\pi t)= p_{\rm vel}(0)/t$ for $n/t \to 0$, with
 $p_{\rm vel} $    
presented in Eq.(\ref{pv0}).

Next we show results where the initial state is Gaussian centered at
the origin
\begin{equation}
\label{Gaussian}
c_n=\frac{1}{\sqrt{S}} e^{-n^2/(4d^2)}~,~~~S=\sum_n e^{-n^2/(2d^2)}
\end{equation} 
where $d$ determines the width
of the discrete Gaussian distribution $c_n^2$.
The numerical evaluation of the corresponding sums
shows that for increasing $d$ the width
of the Gaussian wave packet on the lattice quickly
 approaches the continuum value
 $\Delta x=d$. Already for $d=1$ the deviation is extremely small.
 The expectation value $\langle \hat v^2 \rangle$ can be calculated
analytically. Using  $n^2+(n+2)^2=2(n+1)^2+2$ in Eq. (\ref{v2voncn})
yields
\begin{equation}
\langle \hat v^2\rangle=(1-e^{-2\tilde d^2})/2~,
\end{equation}
with $\tilde d=1/(2d) $.
 The convergence to the continuum value $\Delta
v=\tilde d $ with increasing $d$ is slower than the approach
 of $\Delta x$ to $d$.

\begin{figure}[h]
\label{fighc}
\centering
\epsfig{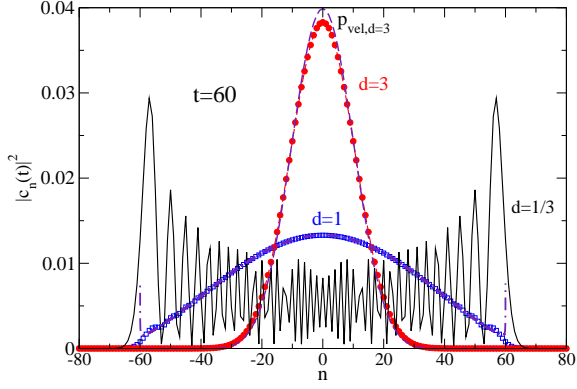}
\vspace{0.5cm}
\caption{Probability to find the particle at site $n$ at time $t=60$
if the initial state was a Gaussian wave packet centered around
the origin: $d=1/3$ (full line), $d=1.0$ (squares)
and $d=3$ (full circles). Also shown is $p_{\rm vel}(n/60)/60$
for $d=1$ (dashed-dotted line)
and $d=3$ (dashed line) with $|n|<60$ treated as a continuous variable.}
\end{figure}

In Fig. 4 we show the time dependent probability  $|c_n(t)|^2$ as
function of $n$ for $t=60$ and three different values of $d$.
In contrast to Fig. 1
only the connecting curves of the values at integer $n$ are shown
for $d=1/3$. 
This probability distribution differs little from the
$t=60$ curve in Fig. 1. The $d=3$ result looks qualitatively 
similar to the broadening of a Gaussian wave packet on the continuum.
The $d=1$ curve shows intermediate behaviour discussed below.

It is again useful to consider the corresponding $p_{\rm vel}(v)$.
 This requires to
calculate $|\langle k|\phi \rangle|^2$, using Eq. (\ref{phivonk}).
For large values of $d$  replacing the $n$-sum
by an integral is a good approximation.
This Gaussian integral
leads to 
\begin{equation}
\label{approx}
|\langle k|\phi \rangle|^2_{\rm appr} =\left (\frac{2d^2}{\pi}\right )^{1/2}
e^{-2d^2k^2}~.
\end{equation} 
Obviously one has lost the $2\pi/a$ periodicity of $|\langle k|\phi
\rangle|^2$, where in this section we have put $a=1$.
This is a minor problem if $|\langle k|\phi \rangle|^2_{\rm appr}$ 
differs from zero in a range smaller than $2\pi$.

 Equation (\ref{pvonv})
 allows to understand the crossover from the result
for $|\phi \rangle=|0\rangle$ to the continuum like curve
for $d=3$ in Fig. 4.
For $v\to \pm 1$ the roots $k_i(v)$ approach the value $\pm \pi/2$. 
Therefore the value of  $|\langle k=\pm \pi/2|\phi \rangle|^2$
is a measure for the suppression of the square root singularities 
of $p_{\rm vel}(v)$. They are almost completely suppressed for $d=3$,
definitely present for $d=1/3$, and $d=1$ is an intermediate case.
In Fig. 4 also $p_{\rm vel}(n/t)/t$ is shown for $d=1$ and $d=3$,
treating $n$ as a continuous variable. For $d=1$ this provides an
excellent approximation except for $|n|$ approaching $60$.
For $d=3$ the deviations for small values of $n$ are larger, as
$\Delta x/\Delta v$ is larger.

\subsection{Wave packets with finite average velocity}

In order for the wave packet to have a finite average velocity
 $\langle \hat v \rangle$ the initial state has to have at least
one pair of neighbouring sites with nonvanishing coefficients $c_n$,
with a least one of them being a complex number.
More generally we consider two-sites states
\begin{equation}
\label{twosites}
|\phi\rangle_m^{(k_0)}=
\frac{1}{\sqrt 2} \left
  ( |0\rangle + e^{imk_0}|m\rangle\right )
\end{equation}
with $m \neq 0$.

Numerical results for $|c_n(t)|^2$ for $m=1$ are shown in Fig. 5. For the
finite velocity case
$k_0=1.0$ the probability distribution 
shows a clear asymmetry with respect to the origin.
As for the single site initial state $p_{\rm vel}(v)$ for  the two
site states can easily be calculated. With
\begin{equation}
|\langle k|\phi\rangle_m^{(k_0)}|^2=\frac{1}{2\pi}\left [1+\cos(m(k-k_0))\right ]
\end{equation}
one obtains using  Eq. (\ref{pvonve}) for $m=1$
\begin{equation}
\label{ptwosites}
p_{\rm vel}(v)=\frac{1+v\sin k_0}{\pi \sqrt{1-v^2}}~.
\end{equation}
For $k_0=\pm \pi/2$ the singularity at the lower (upper) edge of
the distribution is suppressed completely.
Fig. 5 also shows $p_{\rm vel}(n/t)/t$ for $k_0=1.0$.
\begin{figure}[h]
\label{fighc}
\centering
\epsfig{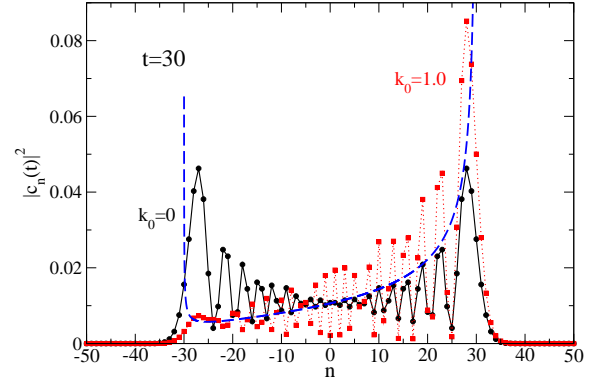}
\vspace{0.5cm}
 \caption{Probability       to find the particle at site $n$ at time $ t$
 if it was localized at the $m=1$ state in Eq. (\ref{twosites})
  at $t=0$ for two different values of $k_0$:
  $k_0=0$ (full line),  $k_0=1$ (dotted line). The dashed line shows
  $p_{\rm vel}(n/t)/t$ for $p_{\rm vel} $  presented 
 in Eq. (\ref{ptwosites}) for $k_0=1.0$. }
\end{figure}

The probability distribution  $p_{\rm vel} $   for $m=2$ is given by
\begin{equation}
p_{\rm vel}(v)=\frac{1}{\pi \sqrt{1-v^2}}\left [1+\cos(2k_0)
-2\cos(2k_0)v^2 \right ]~.
\end{equation}
For $k_0=0$ it reduces to $p_{\rm vel}(v)=2\sqrt{ 1-v^2}/\pi$, i.e.
both singularities are suppressed as shown in Fig. 6.
This example would also have fitted to section IV.A as
the coefficients $c_n$ are real and
$\langle v \rangle =0$.
\begin{figure}[h]
\label{fighc}
\centering
\epsfig{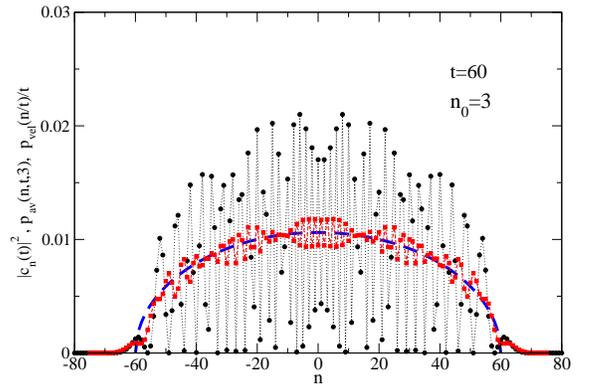}
\vspace{0.5cm}
 \caption{Probability to find the particle at site $n$ at time $ t=60$
   for the $m=2,k_0=0 $
   initial state in Eq. (\ref{twosites})
 (circles). The squares show  $p_{\rm av}(n,t;3)$
and the dashed curve $p_{\rm vel}(n/t)/t $.}
\end{figure}

Next we discuss  boosted  Gaussian initials
state as in Eq. (\ref{boosted}) 
\begin{equation}
  \label{complexG}
c^{(k_0)}_n=e^{ik_0n}\frac{e^{-n^2/(4d^2)}}{\sqrt{S}} ~.
\end{equation}
The expectation value $\langle \hat v^2\rangle^{(k_0)} $
can again be calculated analytically using 
 $n^2+(n+2)^2=2(n+1)^2+2$ in Eq.  (\ref{v2voncn})
 \begin{equation}
   \label{v2k0}
 \langle \hat v^2\rangle^{(k_0)} =(1- \cos 2k_0 e^{-2\tilde d^2})/2.
\end{equation}
For $d$ larger than the lattice constant it is a good approximation
to replace the summation in 
Eq. (\ref{vvoncn}) by an
integration.
Using $n^2 +(n+1)^2
=2(n+1/2)^2+1/2$ one obtains
\begin{equation}
\label{vk0}
\langle \hat v\rangle^{(k_0)}\approx \sin k_0e^{-\tilde d^2/2},~~
\end{equation}
 The velocity uncertainty follows as
\begin{equation}
  \label{vuncert}
(\Delta v)^{(k_0)} \approx \sqrt{(1+e^{-\tilde d^2}\cos 2k_0)(1-e^{-\tilde d^2})/2}~, 
\end{equation}
which explicitely shows the $k_0$ dependence. For large $d$ this
uncertainty reaches the continuum value $\tilde d=1/(2d)$
for $k_0=0$.
For $k_0=\pi/2$ it is proportional to $\tilde d^2$, i.e. the uncertainty
product $\Delta x\Delta v$ vanishes like $1/d$. This implies that also for very
broad wave packets the deviations from the Heisenberg uncertainty product
can be large when the average velocity approaches the maximum velocity on the
lattice as shown in Fig. 7

\begin{figure}[h]
\label{fighc}
\centering
\epsfig{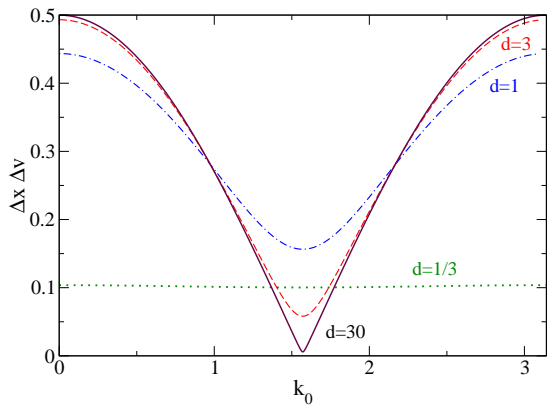}
\vspace{0.5cm}
\caption{Uncertainty product  $\Delta x\Delta v$ as a function of $k_0$
  for different values of $d$: $d=1/3$ (dotted line),
  $d=1$ (dashed-dotted line),   $d=3$ (dashed line), $d=30$ (full line). } 
\end{figure}

For symmetry reasons it is sufficient to consider 
positive values of $k_0$ in the first Brillouin zone. Due to the
properties of the functions $g_n(t)$ mentioned after
Eq. (\ref{cBessel}) the probabilities $|c_n(t)|^2$ are identical
for $k_0$ and $\pi -k_0$. Therefore the largest value we
consider is $k_0=\pi/2$.
    
The corresponding amplitudes $\langle k|\phi \rangle$ follow from
Eq. (\ref{phivonk}) by a shift of the $k$-value.
\begin{equation}
 \langle k|\phi \rangle_{k_0}= \langle k-k_0|\phi \rangle_{k_0=0}~.
\end{equation} 
For $k_0>0$ this leads to $|\langle -\pi/2|\phi \rangle_{k_0} |^2<
|\langle \pi/2|\phi \rangle_{k_0} |^2$. Therefore the singularity
of $p_{\rm vel}(v)$ at $v=-1$ is stronger suppressed than the one
at $v=1$, like in Eq. (\ref{twosites}) for the two-site example.

In Fig. 8
we show results for $k_0=\pi/4, t=60$ and the same three values of $d$ as in 
Fig. 4. The average velocities are  $0.141,0.604$ and $0.676$
for $d=1/3,1$ and $3$.
Except for  a slight asymmetry the $d=3$ curve  
again looks similar to the continuum case.
\begin{figure}[h]
\label{fighc}
\centering
\epsfig{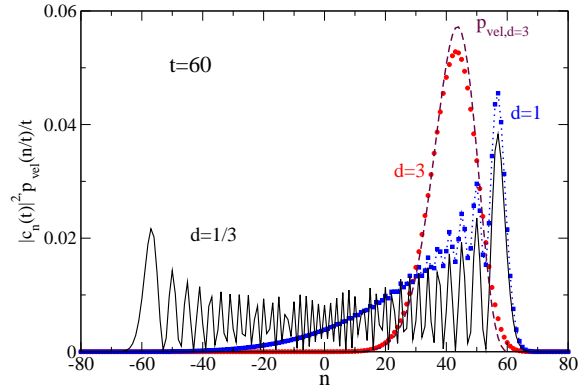}
\vspace{0.5cm}
\caption{Probability to find the particle at site $n$ at time $t=60$
for the complex Gaussian initial state 
with $k_0=\pi/4$ in Eq. (\ref{complexG}): $d=1/3$ (full line),
$d=1.0$ (dashed dotted line) and $d=3$ (full circles).
Also shown for $d=3$ is the  corresponding $p_{\rm vel}(n/60)/60$ (dashed line).} 
\end{figure}
In Fig. 9 the exact results for $d=1$ 
are compared to the corresponding $p_{\rm vel}(n/60)/60$.

\begin{figure}[h]
\label{fighc}
\centering
\epsfig{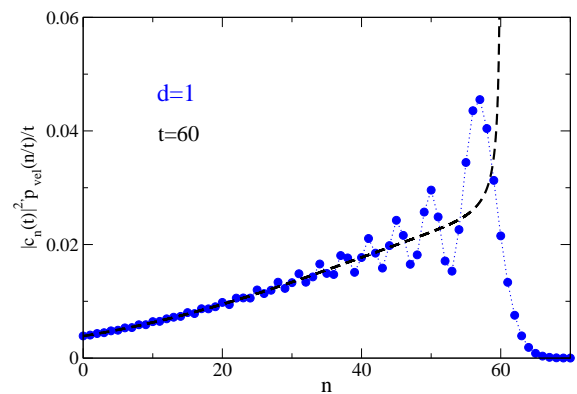}
\vspace{0.5cm}
\caption{Probability to find the particle at site $n$ at time $t=60$
for the complex Gaussian initial state 
with   $d=1$ and $k_0=\pi/4$ in Eq. (\ref{complexG}):
exact results (full circles) and the corresponding
$p_{\rm vel}(n/60)/60$ (dashed line).}

\end{figure}

The value of $|c_n(t)|^2$ for $n=60$ as a function of time is shown
in Fig. 10 for the complex Gaussian initial state 
with $k_0=\pi/4$ for the same values of $d$
as in Fig. 8. 

 \begin{figure}[h]
\label{fighc}
\centering
\epsfig{file=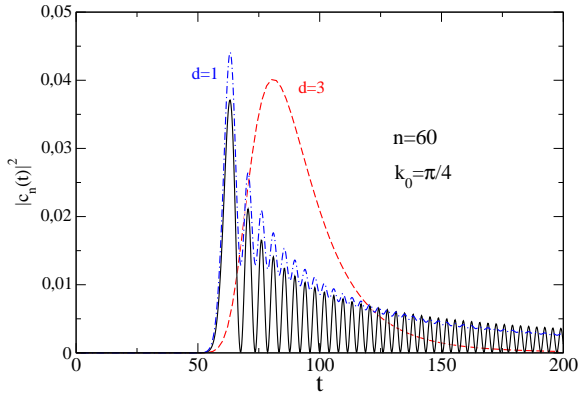,scale=.3,angle=-90}
\vspace{0.5cm}
\caption{  Probability to find the particle at site $n=60$
as a function of time $t$,
if it was in a complex Gaussian initial state of width $d$ and
$k_0=\pi/4$: $d=1/3$ (full
line), $d=1$ (dashed-dotted line) and $d=3$ (dashed line).} 
\end{figure}

For all three values of $d$ the probability to find the particle at
site $n=60$ for times smaller than $t=n/v_{\rm max}=60$ is very small.

We finally consider the case $k_0=\pi/2$, which is very special.
For all positive values of $k_0$ smaller than $\pi/2$ it is
possible to suppress the square root singularity at $v=1$ in
$p_{\rm vel}(v)$ by choosing $d$ large enough.
 For  $k_0=\pi/2$ the maximum
of $|\langle k|\phi \rangle|^2$ is right at   $k=\pi/2$.
The average velocity
takes the largest value. For $d\to \infty$ it
reaches the value $v_{\rm max}=1$.  For $d=3$ and $k_0=\pi/2$
the average velocity is $\langle \hat v \rangle=0.986$.

 As $\epsilon_{\pi/2+\tilde k}=\sin
\tilde k$ the energy dispersion 
is approximately linear around  $k_0=\pi/2$
and $v_{\pi/2+\tilde k}\approx 1-\tilde k^2/2$. It is shown in
Fig. 11 that the deviations from a strictly linear energy dispersion 
become important for large enough times. 

\begin{figure}[h]
\centering
\epsfig{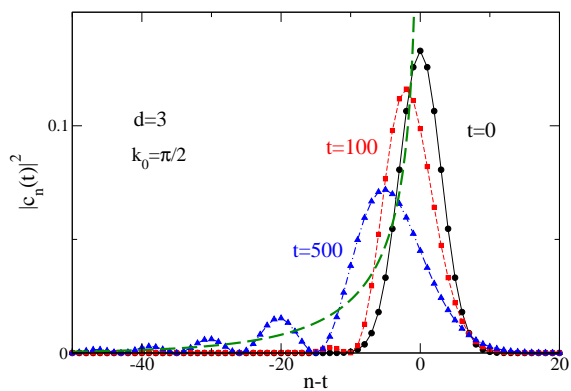}
\vspace{0.5cm}
\caption{Probability distribution $|c_n(t)|^2$ for $k_0=\pi/2$
and $d=3$ as a function of $n-t$ for different times
: $t=0$ circles (full line),
$t=100$ squares (dashed line) and $t=500$ diamonds (dashed-dotted line).
For $t=500$ also $p_{\rm vel}(n/t)/t$  is shown (dashed line).} 
\end{figure}

As the results are shown as a function of $n-t$ there is a visible
shift of the distribution to smaller values for increasing time.
The oscillatory behaviour clearly visible for $n-t<15$ for the $t=500$
distribution is an effect 
that does not occur in the continuum case for a strictly linear
dispersion. It is smoothed out
in the  velocity distribution function, for $d\gg 1$
and $v$ approaching $1$ given by
\begin{equation}
  p_{\rm vel}(v) \approx 2\sqrt{\frac{2d^2}{\pi}}
    \frac{\exp{[-4d^2(1-v)]}}{\sqrt{1-v^2}} ~.
\end{equation}
It is also shown in Fig. 11.

\section{Summary}
Several aspects of the propagation of wave packets in a
one-dimensional tight-binding
chain with nearest neighbour hopping were presented. For an initial
state localized at a single site a totally different time dependence
shows up compared to the well known behaviour of a narrow Gaussian
wave packet in the continuum. An important quantity for the
understanding of the large probabilities to find the particle at
time $t$ near positions $\pm v_{\rm max}t$ is
the divergence of the probability
distribution $p_{\rm vel}(v)$ at the values $\pm v_{\rm max}$.
For strongly localized initial states
the probabilities
$|c_n(t)|^2$ to find the particle at site $n$ is strongly varying with $n$.
Introducing  an appropriate local averaging procedure one can obtain
smoother functions
of $n$.
In the long-time limit these functions
approach
$p_{\rm vel}(n/t)/t$.

In order for the wave packet to have a finite average velocity
at least a two-site initial state involving  a complex amplitude is
necessary. For an initial state with symmetric
nearest neighbour site
occupancies this leads at finite times to an asymmetry in the
site probability distribution. This can easily be understood
 by again analytically calculating  $p_{\rm vel}$. 

Gaussian initial states with zero and finite average velocity
were studied in detail.  Even for initial widths much larger
than the lattice constant  clear differences to the 
continuum case arise for average velocities close to $\pm v_{\rm max}$.

The generalization to higher dimensional lattices is straightforward.     
For a single site initial state in the origin of a square lattice with nearest
neighbour hopping the probability to find the particle at position
$n_x,n_y$ is given by $J_{n_x}(t)^2 J_{n_y}(t)^2$ if the hopping
matrix elements in the $y$-direction are identical to the one
in the $x$-direction.

\section{Acknowledgements}

The author would like to thank Gerhard Hegerfeldt, Reiner Kree,
Salvatore Manmana  and Volker Meden  for a critical
reading of the manuscript and useful comments.

\begin{appendix}

\section{Bessel functions of integer order}

In this appendix we present a simple new way to calculate 
$g_n(t)=\langle n|e^{-i\hat Ht/\hbar}|0\rangle $ which directly shows that it is
proportional to the Bessel function $J_n(\tilde t)$.

The starting point is to write the time evolution
operator $e^{-i\hat Ht/\hbar}$ in product form using 
$\hat H=\epsilon_{01}(\hat T_{\pm a}+ T^\dagger_{\pm  a})$. As the translation
operators $\hat T_a$ and $\hat T_{-a}$ commute one has using Eq. (\ref{BH})
\begin{equation}
e^{-i\hat Ht/\hbar}=e^{i\tilde t \hat T^\dagger_{\pm  a}/2}
e^{i\tilde t \hat T_{\pm  a}/2}
\end{equation}
with $\tilde t =-2 \epsilon_{01}t/\hbar$. This allows a simple calculation
of  $g_n(t)$ without the use of the eigenstates of the Hamiltonian.

For $n\ge 0$ one obtains applying the second factor to the right and
the first one to left  
\begin{eqnarray}
e^{i\tilde t \hat T_{\pm a}/2}|0\rangle &=& \sum_{l=0}^\infty \frac{(i
  \tilde t/2)^l}{l!}|\pm l \rangle \nonumber \\
\langle \pm n|e^{i\tilde t \hat T^\dagger_{\pm a}/2}&=&\sum_{m=0}^\infty
    \langle \pm (n+m)|\frac{(i \tilde t/2)^m}{m!}~.
\end{eqnarray}
Taking the scalar product yields the power series
\begin{equation}
\label{gn}
g_{\pm n}(t)=i^n\left (\frac{\tilde t}{2}\right )^n\sum_{m=0}^\infty
\frac{(-\tilde t^2/4)^m}{(m+n)!m!}=i^nJ_n(\tilde t)~.
\end{equation}
The power series for $J_n(x)$ is usually obtained by solving the Bessel
differential equation by a power series ansatz. \cite{AS}

Alternatively the power series for $g_{\pm n}(t)$ can be obtained by 
performing the $\tilde k$ integration in Eq. (\ref{cBessel}) after
expanding $e^{i \tilde t \cos \tilde k}$ in the integrand. 

The short time approximation for $g_{\pm n}(t)$ follows from 
Eq. (\ref{gn}) as
\begin{equation}
 \label{shorttime} 
g_{\pm n}(t)=i^n \frac{(t/2)^n}{n!}\left (1-\frac{t^2/4}{n+1} +...\right
)~.
\end{equation}
The long time approximation for $J_n(t)$ can be obtained using the
stationary phase method for the integral in  Eq. (\ref{cBessel}).
 For fixed $n$ this yields \cite{AS}
\begin{equation}
J_n(t) \to \sqrt{\frac{2}{\pi t}}
\cos {\left (t-\frac{\pi}{4}-\frac{n\pi}{2}\right )}~.
\end{equation}
A detailed discussion of  $J_n(t)$ for large $n$ and $t \approx n$
can be found in reference 12.
\end{appendix}

\end{document}